\documentstyle[aps,prl,multicol]{revtex}

\def\vec#1{{\bf #1}}

\def\Tr#1{\mbox{Tr}\left\{ #1 \right\} }
\def\sTr#1{\mbox{\small Tr}\left\{ #1 \right\} }

\def\bm#1{{\mbox{\boldmath{$#1$}}}}

\def\strutt{\rule{0pt}{15pt}}
\def\onecol{\vspace*{-18pt} \noindent  \rule{3.4in}{.5pt} \vspace*{3pt}}

\begin{document}

\title{Eigenvalue Density of Correlated Complex Random Wishart Matrices}

\author{Steven H. Simon and Aris L. Moustakas}

\address{Lucent Technologies, Bell Labs, Murray Hill, NJ, 07974}

\date{\today}
\maketitle

\begin{abstract}
Using a character expansion method, we calculate exactly the
eigenvalue density of random matrices of the form $\bf M^\dagger
M$ where $\bf M$ is a complex matrix drawn from a normalized
distribution $P({\bf M}) \sim \exp(-\Tr{\bf A M B M^\dagger})$
with $\bf A$ and $\bf B$ positive definite (square) matrices of
arbitrary dimensions. Such so-called ``correlated Wishart
matrices'' occur in many fields ranging from information theory to
multivariate analysis.
\end{abstract}

\begin{multicols}{2}

Physicists usually think of Wigner and Dyson as the fathers of
random matrix theory\cite{Reviews,JPhys}.  However, twenty years
before their first work on the subject, Wishart\cite{Wishart}
examined random matrices of the form $\bf M M^\dagger$ as a tool
for studying multivariate data.  The properties of these so-called
Wishart Matrices, which are viewed as ``fundamental to
multivariate statistical analysis"\cite{Anderson}, also find
important applications in fields from information theory and
communication\cite{IEEE,Aris,Been}, to mesoscopics\cite{Meso}, to
high energy physics\cite{Janik}, to econo-physics\cite{Stanley}.

In many cases one is interested in Wishart matrices where the
elements of $\bf M$ are not completely independent random
variables, but have correlations along rows and/or columns.
Important examples of this case occur in data analysis
problems\cite{Sengupta}, where random samples have temporal and
spatial correlations, and particularly in wireless communication
and information theory\cite{IEEE,Aris}. The purpose of this paper
is to derive the eigenvalue density of correlated complex Wishart
matrices exactly. A forthcoming longer paper will give more
details of the derivation as well as discussing certain
applications in depth. This problem has previously been studied in
the limit of large matrices where perturbative expansions in $1/N$
can be quite effective\cite{Sengupta,Aris}. If either $\bf A$ or
$\bf B$ is proportional to unity, simpler techniques can be
used\cite{Smith}.

 We first define the problem more precisely.
Let $\bf M$ be an $N$ by $N'$ complex matrix chosen from a
normalized distribution
\begin{equation} P({\bf M}) = \pi^{-N N'} {\cal N}
\exp(-\Tr{\bf A M B M^\dagger})
\end{equation} with $\bf A$ and
$\bf B$ positive definite square matrices which define the
correlations, and Tr indicates trace.  Here, ${\cal N}^{-1} =
\det[{\bf A}]^{N'} \det[{\bf B}]^{N}$ and the factors of $\pi$,
are  normalization constants. An equivalent definition would be to
let ${\bf M = A}^{-1/2} {\bf Z \, B}^{-1/2}$ where $\bf Z$ is a
random complex matrix with independent entries of zero mean and
unit covariance. Note that $\bf A$ is $N$ by $N$ and $\bf B$ is
$N'$ by $N'$. Without loss of generality, we can assume $N \geq
N'$.  For any operator $O({\bf M})$ we define the expectation
bracket $\langle O \rangle$ to be an average over realizations of
$\vec M$ so that $\langle O \rangle \equiv \int d{\bf M} \cdot
O({\bf M}) P({\bf M})$. Note that the normalization is such that $
\langle 1 \rangle=1$.

Let $\lambda_n$ be the $N'$ eigenvalues of $\bf M^\dagger M$ or
equivalently the $N'$ nonzero eigenvalues of $\bf M M^\dagger$ (we
will also have $N - N'$ eigenvalues of $\bf M M^\dagger$ precisely
zero). We define the following quantities to calculate:
\begin{eqnarray}
\label{eq:Gdef} G_\nu(z) &=&   \langle \mbox{$\prod_{n=1}^{N'}$}
(\lambda_n-z)^\nu \rangle = \langle \det  ({\bf M^\dagger
M}-z)^\nu
    \rangle \\ \label{eq:Hdef}
H(z) &=& \strutt \partial G_\nu(z)/\partial \nu|_{\nu=0} =
\langle \mbox{$\sum_{n=1}^{N'}$} \log(\lambda_n-z) \rangle  \\
C(\lambda)\label{eq:Cdef} &=& \strutt \mbox{$\lim_{\epsilon
\rightarrow 0}$} \left[H(\lambda -i 
\epsilon) - H(\lambda + i 
\epsilon) \right]/ 2 \pi i \label{eq:th}  \\
&=& \label{eq:fo} \langle \mbox{$\sum_{n=1}^{N'}$}
\theta(\lambda-\lambda_n) \rangle =
\mbox{$\int_{-\infty}^\lambda$}  dx \, \rho(x)
\\ \rho(\lambda) &=& \strutt dC(\lambda)/d\lambda =
\langle \mbox{$\sum_{n=1}^{N'}$}\delta(\lambda-\lambda_n) \rangle
\label{eq:rhodef}
\end{eqnarray}
where $\theta$ is the step function, $\lambda$ is assumed real,
and in going from Eq. \ref{eq:th} to \ref{eq:fo} we have used
$\lim_{\epsilon\rightarrow 0} \mbox{Im} \log(-y+i\epsilon) =
\pi\theta(y)$
which is true for real $y$.   The quantity we are most interested
in is the eigenvalue density $\rho(\lambda)$.  From Eqs. 
\ref{eq:Gdef}-\ref{eq:rhodef} it is
clear that we can obtain
$\rho$ by calculating $G_\nu(z)$.

Below, we will show
\begin{eqnarray}
\label{eq:mainresult}
    G_\nu(z) &=& Q_\nu(z) R_\nu \,  \det L_{ij} \\
\label{eq:Q}    Q_\nu(z)^{-1} &=& \strutt \Delta_N(a)
     \Delta_{N'}(b)(-z)^{N'(N'-1)/2} J_\nu \\ \label{eq:Jdef}
     J_\nu&=& \strutt \mbox{$\prod_{i=1}^{N-1}$} (\nu + i)^{i} \\
    R_\nu &=& \strutt {\mbox{$\prod_{j=1}^{N - N'-1}$} (N + \nu -j)^{N -
    N'-j}}
\end{eqnarray}
where $R_\nu$ is defined to be unity for $N' \geq N-1$.  In Eq.
\ref{eq:mainresult}, $L_{ij}$ is an $N$ by $N$ matrix with
elements $L_{ij} = g(a_i b_j;\nu+N,z)$ for $j \leq N'$ and $L_{ij}
= a_i^{j-1}$ for $j > N'$ where we have defined $a_i$ and $b_j$ to
be the eigenvalues of the matrices $\bf A$ and $\bf B$.  The
function $g$ is given by
\begin{eqnarray}
    g(x;\alpha,z) &=& x^{N-\alpha} e^{-z x} \Gamma(\alpha,-z x)
    \label{eq:g} \\
    &=& \strutt x^N \mbox{$\int_0^\infty d\lambda$} (\lambda - z)^{\alpha-1} e^{-x
    \lambda} \label{eq:g3}
\end{eqnarray}
with $\Gamma$ the incomplete gamma function\cite{Gradstein}, and
we note that for integer $\alpha > 0$ we have the simple form
\begin{equation}
\label{eq:g2}
    g(x,\alpha,z) = x^{N-\alpha} (\alpha - 1)! \,
    \mbox{$\sum_{m=0}^{\alpha-1}$} (-z x)^m / m!
\end{equation}
In Eq. \ref{eq:Q} and throughout this paper we use the notation
\begin{equation}
\label{eq:vandef}
    \Delta_V(x) = \prod_{1 \leq i < j \leq V} (x_j - x_i) =
    \det[x_j^{i-1}]
\end{equation}
to represent a $V$-dimensional Vandermonde determinant.

From $G_\nu(z)$ we calculate $C(\lambda)$ using Eqs. \ref{eq:Hdef}
and \ref{eq:Cdef}.  The differentiation (Eq. \ref{eq:Hdef}) with
respect to $\nu$ brings down a factor of $\log(\lambda - z)$ in
the argument of Eq. \ref{eq:g3}.  This log becomes a step function
when the limit is taken in Eq.
\ref{eq:Cdef}. We obtain
\begin{equation}
\label{eq:cans}
    C(\lambda) = N'- Q_0(\lambda) R_0\sum_{n=1}^N \det K^{(n)}_{ij}
\end{equation}
where we also used
$G_{\nu=0}(z)\equiv 1$.
 Here, we have defined $N$ by $N$ matrices
$K^{(n)}$  with $K^{(n)}_{ij} = g(a_i b_j; N,\lambda)$ for $n \neq
i$ and $j \leq N'$ and $K^{(n)}_{ij} = a_i^{j-1}$ for $j
> N'$ and $n \neq i$. For the case of $i=n$ we have  $K^{(n)}_{nj}
= e^{-a_n b_j \lambda} (N-1)! $ for $j \leq N'$ and $K^{(n)}_{nj}
= 0$ for $j > N'$.

We then differentiate Eq. \ref{eq:cans}  (see Eq. \ref{eq:rhodef})
to obtain
\begin{equation}
    \rho(\lambda)= Q_0(\lambda) R_0\sum_{n=1}^N \left[ \det \tilde K^{(n)}_{ij} +
    \!\!\!\!
    \sum_{m=1, m \neq n}^N  \!\!\!\! \det T^{(nm)}_{ij} \right]
\end{equation}
where  $\tilde K^{(n)}$ and $T^{(nm)}$ are $N$ by $N$ matrices
with elements defined as follows: $\tilde
K^{(n)}_{ij}=K^{(n)}_{ij}$ for $i \neq n$ and $\tilde K^{(n)}_{nj}
= [a_n b_j + (N'(N'-1)/2)/\lambda] K^{(n)}_{nj}$ for $j \leq N'$
and $\tilde K^{(n)}_{nj} = 0$ for $j
> N'$. Also $T^{(nm)}_{ij} = K^{(n)}_{ij}$ for $i \neq
m$ with $T^{(nm)}_{mj} = (N-1) a_m b_j\, g(a_m b_j,N-1,\lambda)$
for $j \leq N'$ and $T^{(nm)}_{mj} = 0$ for $j > N'$.

The above expressions are our main results. The remainder of this
paper comprises the proof of Eqs. \ref{eq:mainresult}-\ref{eq:g}
from which all of our other results follow. We start by focusing
on the case of square matrices $\bf M$ (so $N = N'$). We write $$
G_\nu(z) = {\cal N}\pi^{-N^2}\!\!  \int \! d{\bf M} \,
e^{-\sTr{\bf A M B M^\dagger}} \, \det ({\bf M M}^\dagger - z)^\nu
$$
we then define $\bf M = U m V$ where $\vec m$ is a diagonal matrix
of the singular values $m_i$
 of $\bf M$ and $\bf U$ and $\bf V$ are unitary
matrices. We then separate the integral over $\bf M$ into
integrals over the eigenvalues $\lambda_i = |m_i|^2$ of $\bf M
M^\dagger$ and ``angular" integrals over $\bf U$ and $\bf V$. This
approach, common in random matrix theory\cite{Reviews}, yields
\begin{equation}
\label{eq:G2} G_\nu(z) = {\cal C N} \int d{\bm \lambda}
\prod_{j=1}^N ( \lambda_j - z)^\nu   \,\, \Delta_N(\lambda)^2
D_{\bf A, B}( {\bm \lambda})
\end{equation}
where ${\bm \lambda} = \bf m m^\dagger$ is the diagonal matrix of
eigenvalues $\lambda_i$ and $\int d{\bm \lambda} = \prod_{i=1}^N
\int_0^\infty d\lambda_i$ and $\cal C$ is an $N$-dependent
numerical constant (which we will not keep track of explicitly but
will fix at the end of the calculation). Here,
$\Delta_N(\lambda)^2$ is the Vandermonde determinant squared of
the $\lambda_i$'s (which is the Jacobian of the transformation)
and
$$
D_{\bf A,B}({\bm \lambda})=  \int_{U(N)}\!\!\!\!\!\! d{\bf U}
\int_{U(N)}\!\!\!\!\!\! d{\bf V} \, e^{-\sTr{\bf A U m V B
V^\dagger m^\dagger U^\dagger}}
$$
where $\bf U$ and $\bf V$ are $N$ by $N$ unitary matrices which
are integrated with the usual Haar measure over $U(N)$.  Note that
we have written the integral $D_{\bf A,B}$ as a function of ${\bm
\lambda} = \bf m m^\dagger$ (we will see this is indeed true).
Here ${\cal C}{\cal N} \Delta_N(\lambda)^2 D_{\bf A, B}({\bm
\lambda})$ is precisely the joint probability density of the
$\lambda$'s. As such, it is clear that this density (and also $D$)
must vanish exponentially if any of the $\lambda$'s is taken to
infinity (this will be important below).
In particular, when ${\bf A}$, $\bf B$ are identity matrices we
can see that $D_{\bf A, B}({\bm \lambda}) \sim \exp(-\sum_i
\lambda_i)$.

To address these integrals over $U(N)$, we use the character
expansion method discussed in depth in Ref. \onlinecite{Bal}. This
allows us to write
$$ e^{-\sTr{\bf A U m V B V^\dagger m^\dagger
U^\dagger}}\! = \! \mbox{$\sum_r$}   \alpha_r \chi_r ({\bf A U m V
B V^\dagger m^\dagger U^\dagger})
$$
where $\alpha_r$ are expansion coefficients (discussed below), the
sum is over representations $r$ of $Gl(N)$, and $\chi_r$ is the
character of the group element in the proper representation. (Note
that the representation theory of $Gl(N)$ is identical to that of
$U(N)$). A character is just the trace taken in the proper
representation, so we have
$$ \chi_r
({\bf A U m V B V^\dagger m^\dagger U^\dagger}) \!=\!\! A^r_{ab}
U^r_{bc} m^r_{cd} V^r_{de} B^r_{ef} V^{r*}_{gf} m^{r*}_{hg}
U^{r*}_{ah}
$$
with lower repeated indices summed (and superscripts $r$ tell us
that the matrix is in representation $r$).   We now use the
orthogonality property\cite{Bal}
$$
 \int_{U(N)}\!\!\!\!\!\!\!\!\!\! d{\bf U}\,\,  U^r_{ab} U^{r*}_{cd} =
 d_r^{-1} \delta_{ac} \delta_{bd}
$$
with $d_r$ the dimension of representation $r$ (discussed below).
Combining the above three equations we obtain
\begin{equation}
\label{eq:D2}
 D_{\bf A, B}({\bm \lambda}) =\mbox{$\sum_r$}
\alpha_r d_r^{-2}\,\, \chi_r({\bf A}) \chi_r({\bf B}) \chi_r({\bm
\lambda})
\end{equation}
 As discussed in Ref. \onlinecite{Bal}, each
representation $r$ is specified by a set of increasing integers $
0  \leq k_N < k_{N-1} \ldots <k_1$, so the sum written over $r$ is
actually a ordered sum over the $k$'s ($k_j = N + n_j - j$ in the
notation of Ref. \onlinecite{Bal}).   In Ref. \onlinecite{Bal} it
it is also found that $ \alpha_r  =   s(k) \det[ 1 / (k_j +i -N)!
] = s(k) \Delta_N(k)/ C(k)$ where $C(k) =  \prod_{j=1}^N k_j!$.
Here $s(k)=(-1)^v$ with $v=N(N-1)/2 - \sum_j k_j$. In the same
work\cite{Bal} it is also shown that $\alpha_r / d_r
    = s(k) F_N / C(k)$ with $F_N = \prod_{j=1}^{N-1} j!$ from which we then obtain
\begin{equation}
\label{eq:ad2}
    \alpha_r d_r^{-2} =  s(k) F_N^2 / \Delta_N(k) C(k)
\end{equation}
The Weyl character formula tells us that\cite{Bal}
\begin{equation} \label{eq:Weyl} \chi_r({\bf X})
=\det[x_i^{k_j}]/\Delta_N(x)
\end{equation}
with $k_j$ the integers describing the representation $r$, and
$x_i$ the eigenvalues of $\bf X$.  Plugging Eq. \ref{eq:Weyl} into
Eq. \ref{eq:D2} we obtain
\begin{eqnarray}
\label{eq:star}  \Delta_N(\lambda)^2 D_{\bf A, B} ( {\bm \lambda})
&=& \Delta_N(\lambda) \mbox{$\sum_r$}  \Phi_r
\det[\lambda_i^{k_j}]
\\
\Phi_r &=& \strutt \alpha_r d_r^{-2}\,\, \chi_r({\bf A})
\chi_r({\bf B})\label{eq:Phidef}
 \end{eqnarray}

We next need a useful identity:
$$
    \Delta_N(\lambda) \!=\!  \frac{1}{(-z)^{N(N-1)/2}}
    \det\!\left[\!\!\left(\frac{\lambda_i}{\lambda_i-z}\right)^{\!j-1}\!\right]
     \!\prod_{n=1}^N
    (\lambda_n-z)^{N-1}
$$
To show this we note that since $\Delta_N(\lambda) = \prod_{i < j}
(\lambda_i - \lambda_j)$ we can freely add a constant to each
$\lambda_i$ and not change $\Delta_N$.  Thus, we have
$\Delta_N(\lambda) = \Delta_N(\lambda - z)$.  Next we use
$\Delta_N(x_1, \ldots, x_N) =  \Delta_N(-1/x_1, \ldots, -1/x_N)
\prod_{j=1}^N x_j^{N-1}$ so that we can relate $\Delta_N(\lambda)$
to $\Delta_N(1/[z-\lambda])$. We then use $1/(z -\lambda)  - 1/z =
\lambda/[z(z-\lambda)]$ and we again shift each term in the
Vandermonde determinant by $-1/z$. Finally we separate out factors
of $-z$  and write the Vandermonde determinant as on the far right
of Eq. \ref{eq:vandef}.

 Using the above expression for $\Delta_N(\lambda)$ and plugging Eq.
\ref{eq:star} into Eq. \ref{eq:G2} yields
\end{multicols}
\onecol
\begin{eqnarray}
 G_\nu(z) &=& \frac{\cal C {\cal N} }{ (-z)^{N(N-1)/2}} \, \int d{\bm \lambda}
\prod_{i=1}^N ( \lambda_i - z)^{\nu+N-1} \sum_r \Phi_r
\det[\lambda_i^{k_j}]
\det\left[\left(\frac{\lambda_i}{\lambda_i-z}\right)^{j-1}\right]
\\
&=& \frac{\cal C {\cal N}}{  (-z)^{N(N-1)/2}} \sum_{c_1, \ldots,
c_N} \epsilon_{c_1 \ldots c_N}  \int d{\bm \lambda} \prod_{i=1}^N
( \lambda_i - z)^{\nu+N-c_i} \left\{ \left[  \sum_r \Phi_r
 \sum_{d_1 \ldots d_N}
\epsilon_{d_1 \ldots d_N} \prod_{i=1}^N \lambda_i^{k_{d_i}}
\right]\prod_{i=1}^N \lambda_i^{c_i-1} \right\}
\label{eq:secondline}
\end{eqnarray}
\vspace*{-.2in}
\begin{multicols}{2}
In Eq. \ref{eq:secondline} we have rewritten the determinants as
sums over all permutations by using the completely antisymmetric
Levi-Cevita tensor $\epsilon_{c_1 \ldots c_N}$ which is 1 if $c_1,
\ldots, c_N$ is an even permutation of $[1, \ldots, N]$, is $-1$
if it is an odd permutation, and is otherwise zero. As mentioned
above, the quantity ($\Delta_N D$) in the square brackets in Eq.
\ref{eq:G2}  is exponentially convergent to zero when any
$\lambda_i$ becomes large (Thus the quantity in the curly brackets
is also exponentially convergent). Further, so long as $c_i-1 \neq
0$ (which implies $c_i -1 + k_{d_i} \neq 0$) the quantity in curly
brackets goes to zero at the lower boundary $\lambda_i = 0$. This
enables us to trivially integrate by parts with respect to
$\lambda_i$ where we differentiate the quantity in the curly
brackets and integrate the quantity outside of the curly brackets
and we do not obtain any boundary terms. We choose to do this
integration exactly $c_i - 1$ times to obtain
\end{multicols}
\onecol
\begin{eqnarray}
\nonumber
 G_\nu(z)
&=& \frac{\cal C {\cal N}}{  (-z)^{N(N-1)/2}} \sum_{c_1, \ldots,
c_N} \!\! \epsilon_{c_1 \ldots c_N} \int d{\bm \lambda}
\prod_{i=1}^N ( \lambda_i - z)^{\nu+N-1}   (-1)^{c_i-1} \sum_r
\Phi_r
 \sum_{d_1 \ldots d_N} \!\!
\epsilon_{d_1 \ldots d_N} \prod_{i=1}^N \lambda_i^{k_{d_i}} \,\,
\prod_{p=1}^{c_i - 1} \frac{k_{d_i} +p}{\nu + N  - p}
\end{eqnarray}
\vspace*{-.2in}
\begin{multicols}{2}
We would now like to interchange the order of integration and
summation such that all integrals are done first.  However, if we
did this we would end up with divergent integrals.  To fix this
problem, we insert a cutoff function such as $f(\lambda) =
\exp[-\delta \lambda]$ and at the end of the calculation we will
take $\delta$ to zero.  (The precise form of the cutoff function
will not matter). This allows us to reorder and write
\end{multicols}
\onecol
\begin{eqnarray}
\nonumber
 G_\nu(z)
&=& \frac{\cal C {\cal N} }{(-z)^{N(N-1)/2}}  \sum_r \Phi_r
\!\!\!\! \sum_{c_1 \ldots c_N} \!\! \epsilon_{c_1 \ldots c_N}
\!\!\!\! \sum_{d_1 \ldots d_N} \!\!  \epsilon_{d_1 \ldots d_N}
 \prod_{i=1}^n  (-1)^{c_i-1} \left[ \int_0^\infty d\lambda_i
 f(\lambda_i) ( \lambda_i - z)^{\nu+N-1}
 \lambda_i^{k_{d_i}}  \prod_{p=1}^{c_i - 1}
\frac{k_{d_i} +p}{\nu + N  - p} \right]
\end{eqnarray}
\vspace*{-.2in}
\begin{multicols}{2}
We can now do the sums over $c$'s and $d$'s to obtain
\begin{eqnarray}
\label{eq:Gshort} G_\nu(z) &=& N! \, \, {\cal C N} \,
z^{-N(N-1)/2} \,\, \mbox{$\sum_r$} \Phi_r \det[P_{ij}^{(r)}]
\\ P^{(r)}_{ij} &=& \strutt P^{(r)}_{i1} \mbox{$\prod_{p=1}^{j - 1}$} \frac{k_{i}
+p}{\nu + N  - p} \label{eq:Pform} \\
P^{(r)}_{i1} &=& \strutt \mbox{$\int_0^\infty d\lambda$} \,
 f(\lambda)\,  ( \lambda - z)^{\nu+N-1}
 \,
 \lambda^{k_{i}}
\end{eqnarray}
The rather special form of the matrix expressed in Eq.
\ref{eq:Pform} allows us to calculate the determinant
straightforwardly yielding $\det[{P}^{(r)}_{ij}] = \Delta_N(k)
J^{-1}_\nu \prod_{i=1}^N P_{i1}^{(r)}$. Thus we have $$  \Phi_r
\det[{P}^{(r)}_{ij}] =   \frac{ F_N^2 \det[a_i^{k_j}]
\det[b_i^{k_j}] \prod_{i=1}^N \frac{(-1)^{k_i}}{k_i!}
P_{i1}^{(r)}}{(-1)^{N(N-1)/2} \Delta_N(a) \Delta_N(b) J_\nu}
$$
where we have used Eq. \ref{eq:Jdef}, \ref{eq:ad2}, \ref{eq:Weyl},
and \ref{eq:Phidef} and the definitions of $C(k)$ and $s(k)$.
Plugging this result into Eq. \ref{eq:Gshort} we now need only do
the sum over $r$. This sum, as explained above, is actually a sum
over $0 \leq k_N < k_{N-1} \ldots < k_1$.  Thus we have
\begin{eqnarray}
G_\nu(z) &=&  S \sum_{0 \leq k_N \ldots < k_1} \det[a_i^{k_j}]
    \det[b_i^{k_j}] \, \mbox{$\prod_{i=1}^N$} w(k_i)  \label{eq:Scauchy}
\\    w(k) &=& \frac{(-1)^k \strutt }{k!} \mbox{$\int_0^\infty d\lambda$} \,
 f(\lambda)\,  ( \lambda - z)^{\nu+N-1}
 \lambda^{k}
\end{eqnarray}
with $S = {\cal N C} Q_\nu(z)$  and we have absorbed the numerical
constants $F_N$ into $\cal C$.  We can now address the sum in Eq.
\ref{eq:Scauchy} using the Cauchy-Binet Theorem (see Appendix) to
obtain $
    G_\nu(z) = S \det[W(a_i b_j)]
$
with the function $W(x)$ defined by
$$
    W(x) = \sum_{k=0}^\infty x^k w(k)  = \mbox{$\int_0^\infty d\lambda$} \,
 f(\lambda)\,  ( \lambda - z)^{\nu+N-1}  e^{-x \lambda}
$$
We can now remove the convergence function $f$ (letting $\delta
\rightarrow 0$ as discussed above) to obtain (See Eq. \ref{eq:g3})
$ W(x)  \rightarrow g(x,\nu+N,z) x^{-N}$. The factors of $x^{-N}$
precisely cancel the prefactor ${\cal N}$ and we recover the
desired result Eq. \ref{eq:mainresult} for $N=N'$ (where
$R_\nu=1$) up to the $N$-dependent normalization prefactor $\cal
C$ which we have not kept track of. To show that the normalization
of Eq. \ref{eq:mainresult} (i.e., ${\cal C}=1$) is indeed correct
we need only verify $G_0(z) = 1$.  To do this (using Eq.
\ref{eq:g2}) we need to establish
$$
    \det\left[(N-1)! \sum_{m=0}^{N-1}  \frac{(-z a_i b_j)^m}{  m!} \right] =
    Q(z,0)^{-1}
$$
which is easy to show using the Cauchy-Binet theorem (In Eq.
\ref{eq:cauchy}, use $w(k) = (-z)^k (N-1)!/k!$ for $k \leq N-1$
and $w(k)=0$ otherwise, so the determinants on the left hand side
of Eq. \ref{eq:cauchy} are precisely $\Delta(a)$ and $\Delta(b)$)
which completes the proof for the case of $N=N'$.

Using the results we have derived for square matrices we can now
easily derive results  for rectangular matrices ($N > N'$). Given
an $N'$ dimensional matrix $\bf B$ with eigenvalues $b_1, \ldots,
b_{N'}$ we consider an auxiliary $N$ dimensional matrix $\bf
\tilde B$ with the $N'$ eigenvalues $b_1, \ldots, b_{N'}$ plus
$N-N'$ eigenvalues $b_{N'+1}, \ldots, b_N$. We then take a limit
where $b_{N'+1}, ..., b_N$ all go to infinity.  By viewing the
matrix $\bf M$ as being ${\bf A}^{-1/2} {\bf Z \, \tilde
B}{}^{-1/2}$ it is clear that taking this limit drives $N-N'$
columns of $\bf M$ to zero and we obtain effectively an $N$ by
$N'$ dimensional problem (with $N-N'$ additional zero
eigenvalues).  To take these limits we will use the expansion
 (Ref. \onlinecite{Gradstein} Eq.
8.357)
\end{multicols}
\onecol
\begin{equation}
\label{eq:gammaexpand}
    \lim_{x \rightarrow \infty} g(x,N+\nu,z) = x^{N-1}
    (-z)^{N + \nu -1} \left[1 + \frac{N + \nu -1}{(-zx)} +
    \frac{(N + \nu-1)(N + \nu -2)}{(-zx)^2} + \ldots \right]
\end{equation}
\begin{multicols}{2}
\vspace*{-.25in} \noindent As $b_N \rightarrow \infty$ we use the
first term of this expansion and replace the $j=N$ row of the
matrix $L_{ij}$ in Eq. \ref{eq:mainresult} with $(a_i b_N)^{N-1}
(-z)^{ N  + \nu -1}$.  In the denominator ($Q_\nu$) we have
$\Delta_{N}(b) \rightarrow (b_N)^{N-1} \Delta_{N-1}(b)$ so the
factors of $b_N$ cancel to give a finite ratio.  We next let
$b_{N-1} \rightarrow \infty$. In taking this limit the first term
in the expansion Eq. \ref{eq:gammaexpand} would result in the
$j=N-1$ row of $L_{ij}$ being exactly proportional to the $j=N$
row and thus we would obtain $\det L_{ij} = 0$.   Thus, the
leading divergence as $b_{N-1} \rightarrow \infty$ is actually
from the second term of the expansion \ref{eq:gammaexpand}.  We
can then replace the $j=N-1$ row of $L_{ij}$ with $(N + \nu -1)
(a_i b_{N-1})^{N-2} (-z)^{N + \nu -2}$.  Again the diverging
powers of $b_{N-1}$ here are cancelled by powers in the
denominator $(Q_\nu)$ since $\Delta_{N-2}(b) = (b_{N-1})^{N-2}
\Delta_{N-1}(b)$.  This procedure can be continued until we have
let all $b_{N'+1} \ldots b_N \rightarrow \infty$.  We cancel all
of the diverging terms then factor out the numerical prefactors
(such as $N + \nu -1$) to give $R_\nu$,  and factor out common
factors of $-z$ to obtain the general result quoted above in Eq.
\ref{eq:mainresult}-\ref{eq:g} times $(-z)^{\nu (N-N')}$ which is
due to the fact that, as mentioned above, our auxiliary problem
has $N-N'$ zero eigenvalues.  In this way we complete our more
general proof.

{\sl Appendix: Cauchy-Binet Theorem}\cite{Bal}.  Given
$N$-dimensional vectors $a_i$ and $b_i$, and a function $W(z) =
\sum_{i=0}^\infty w(i) z^i$ convergent for $|z| < \rho$ then if
$|a_i b_j| < \rho$ for all $i,j$ we have (where the determinants
are all taken with respect to the indices $i$ and $j$):
\begin{equation}
\label{eq:cauchy} \sum_{0 \leq k_N \ldots < k_1}  \!\!\!\!\!
\det[a_i^{k_j}]
    \det[b_i^{k_j}]\, \mbox{$\prod_{i=1}^N$} w(k_i) = \det[W(a_i b_j)]
\end{equation}

\end{multicols}

\end{document}